# ON THE AMBAINIS-BACH-NAYAK-VISHWANATH-WATROUS CONJECTURE

**CLEMENT AMPADU**

College of Professional Studies
Northeastern University
360 Huntington Avenue
Boston, Massachusetts, 02115
U. S. A.
e-mail: drampadu@hotmail.com

**Abstract**

We show the flaw in a theorem of Konno-Namiki-Soshi-Sudbury in [3], provide the necessary correction in the case of the finite Hadamard walk and use it to show that a conjecture of Ambainis-Bach-Nayak-Vishwanath-Watrous in [1] is false.

## 1. Introduction

Ambainis et al. [1] defined and analyzed quantum computational variants of random walks on one-dimensional lattices. Of particular interest is their analysis of the Hadamard walk which is a quantum analog of the symmetric random walk. Ambainis et al. [1] and Konno et al. [3] have formally defined the notion of the quantum walk as well as the Hadamard walk itself, albeit somewhat slightly different notations. They both considered absorption problems for the Hadamard walk on the sets $\{0, 1, ..., N\}$ and $\{0, 1, ...,\}$ which corresponds to the *finite Hadamard* walk and the *semi-infinite Hadamard* walk, respectively. In the finite case, we have absorbing barriers at 0 and *N*, whilst in the semi-infinite case we have the absorbing barrier at 0. The particular interest of Ambainis et al. [1] was the answer to the following question: *What are the absorption probabilities of the walks*?

Let

$$\Phi = \left\{ \varphi = \begin{bmatrix} \alpha \\ \beta \end{bmatrix} \in C^2 : |\alpha|^2 + |\beta|^2 = 1 \right\}$$

be the collection of initial quibit states of which the absorption probabilities depend upon. For the semi-infinite Hadamard walk, let $P_1^\infty(\varphi)$ be the probability that the particle hits zero starting from location 1, and for the finite Hadamard walk, let $P_1^N(\varphi)$ be the probability that the particle hits zero starting from location 1 before it arrives at location $N$. For the semi-infinite walk, Ambainis et al. [1] have shown that $P_1^\infty({}^t[0, 1]) = \frac{2}{\pi}$, whilst Konno et al. [3] gave the following generalization $P_1^\infty(\varphi) = \frac{2}{\pi} + 2\left(1 - \frac{2}{\pi}\right)\text{Re}(\overline{\alpha}\beta)$ for any initial quibit state $\varphi \in \Phi$. For the finite Hadamard walk, Ambainis et al. [1] studied the asymptotic behavior of $P_1^N({}^t[0, 1])$ and concluded $P_1^N({}^t[0, 1]) = \frac{1}{\sqrt{2}}$ as $N \to \infty$. Unable to derive a closed form solution for $P_1^N({}^t[0, 1])$, they conjectured that $P_1^N({}^t[0, 1])$ obeys the following recursive formula:

$$P_1^{(N+1)}({}^t[0, 1]) = \frac{1 + 2P_1^N({}^t[0, 1])}{2 + 2P_1^N({}^t[0, 1])}; \quad N \geq 1,$$

$$P_1^1({}^t[0, 1]) = 0.$$

This recursive formula is Conjecture 11 in [1] stated using the notation of Konno et al. [3]. The objective of this paper is to show that $P_1^N({}^t[0, 1])$ does not obey the above recursive formula. In particular, we will show that the conjecture is false in the case $N = 3$. Let us remark that the conjecture asserts that $P_1^3({}^t[0, 1]) = \frac{2}{3}$. This paper is organized as follows: in Section 2, we show that Theorem 2 in [3] in the present state is flawed, and give the necessary correction for the finite Hadamard walk. In Section 3, we use the correction to show that $P_1^N({}^t[0, 1])$ does not obey the recursive formula as mentioned earlier. In particular, we show $P_1^3({}^t[0, 1]) \neq \frac{2}{3}$. In Section 4, we leave the reader with some concluding remarks and further open problems.

## 2. The Absorption Probability for the Finite Hadamard Walk

Recall that the time evolution of the one-dimensional quantum walk is given by the following unitary matrix $U = \begin{bmatrix} a & b \\ c & d \end{bmatrix}$, where $a$, $b$, $c$ and $d$ are complex numbers. The unitary matrix of the Hadamard walk is defined by

$$U = H = \begin{bmatrix} \frac{1}{\sqrt{2}} & \frac{1}{\sqrt{2}} \\ \frac{1}{\sqrt{2}} & \frac{-1}{\sqrt{2}} \end{bmatrix}.$$

### 2.1. The flaw

If $U \neq H$, then Konno et al. [3] gave a formula for $P_k^N(\varphi)$ in Theorem 2 of their paper which can be interpreted as the absorption probability for the finite quantum walk in one-dimension irrespective of the initial location $k \in \{1, 2, ..., N-1\}$ and the initial quibit state $\varphi \in \Phi$ of the particle. However, we note from Theorem 2 of their paper that $p_k^N(z)$ and $r_k^N(z)$ are supposed to satisfy

$$p_k^N(z) = \left(\frac{z}{2} + E_z\right)\lambda_+^{k-1} + \left(\frac{z}{2} - E_z\right)\lambda_-^{k-1}, \qquad (2.1.1)$$

$$r_k^N(z) = C_z(\lambda_+^{k-N+1} - \lambda_-^{k-N+1}), \qquad (2.1.2)$$

where $\lambda_\pm = \dfrac{z^2 - 1 \pm \sqrt{z^4 + 1}}{\sqrt{2}z}$ and $C_z$ and $E_z$ are given by

$$C_z$$

$$= \frac{z^2}{\sqrt{2}}(-1)^{N-2}(\lambda_+^{N-3} - \lambda_-^{N-3})$$

$$\times \{(\lambda_+^{N-2} - \lambda_-^{N-2})^2 - \frac{z}{\sqrt{2}}(\lambda_+^{N-2} - \lambda_-^{N-2})(\lambda_+^{N-3} - \lambda_-^{N-3}) - (-1)^{N-3}(\lambda_+ - \lambda_-)^2\}^{-1}$$

$$(2.1.3)$$

and

$$E_z = \frac{z}{2(\lambda_+^{N-2} - \lambda_-^{N-2})}$$

$$\times \left[ \left( \begin{array}{l} 2(-1)^{N-3}(\lambda_+ - \lambda_-)(\lambda_+^{N-3} - \lambda_-^{N-3}) \\ \left\{ \begin{array}{l} (\lambda_+^{N-2} - \lambda_-^{N-2})^2 \\ -\frac{z}{\sqrt{2}}(\lambda_+^{N-2} - \lambda_-^{N-2})(\lambda_+^{N-3} - \lambda_-^{N-3}) - (-1)^{N-3}(\lambda_+ - \lambda_-)^2 \end{array} \right\}^{-1} \\ + (\lambda_+^{N-2} + \lambda_-^{N-2}) \end{array} \right) \right]. \quad (2.1.4)$$

In the case $N = 3$, equations (2.1.2) and (2.1.3) show that $r_1^3(z) = 0$. However, this is contrary to the boundary condition $r_{N-1}^N(z) = 0$ which evidently shows that $r_1^N(z) = 0$ if and only if $N = 2$.

### 2.2. The correction for the finite Hadamard walk

To give the correction in the case of the finite hadamard walk, we begin with the following:

**Lemma 2.2.1.** $p_k^N(z)$ and $r_k^N(z)$ satisfy

$$p_k^N(z) = \frac{z(\lambda_-)^N}{(\lambda_-)^N - (\lambda_+)^N} \lambda_+^{k-1} + \frac{z(\lambda_+)^N}{(\lambda_+)^N - (\lambda_-)^N} \lambda_-^{k-1}$$

and

$$r_k^N(z) = \frac{z(\lambda_-)^N}{(\lambda_-)^N - (\lambda_+)^N} \lambda_+^{k+1} + \frac{z(\lambda_+)^N}{(\lambda_+)^N - (\lambda_-)^N} \lambda_-^{k+1}.$$

**Proof.** Konno et al. [3] have shown that

$$p_k^N(z) = azp_{k-1}^N(z) + czr_{k-1}^N(z) \quad \text{and} \quad r_k^N(z) = bzp_{k+1}^N(z) + dzr_{k+1}^N(z).$$

If $U = H$ (the Hadamard walk), $p_k = (\lambda_+)^k$ and $r_k = (\lambda_-)^k$, then

$$p_k^N(z) = \frac{z}{\sqrt{2}}(\lambda_+)^{k-1} + \frac{z}{\sqrt{2}}(\lambda_-)^{k-1}$$

and

$$r_k^N(z) = \frac{z}{\sqrt{2}}(\lambda_+)^{k+1} - \frac{z}{\sqrt{2}}(\lambda_-)^{k+1}.$$

However, Konno et al. [3] have shown that the boundary conditions for the finite quantum walk in one-dimension are given by $r_{N-1}^N(z) = 0$ and $p_1^N(z) = z$. Moreover, $p_k^N(z) = \frac{z}{\sqrt{2}}(\lambda_+)^{k-1} + \frac{z}{\sqrt{2}}(\lambda_-)^{k-1}$ and $r_k^N(z) = \frac{z}{\sqrt{2}}(\lambda_+)^{k+1} - \frac{z}{\sqrt{2}}(\lambda_-)^{k+1}$ do not satisfy the boundary conditions. However, the identical nature of the coefficients of $(\lambda_\pm)^{k-1}$ and $(\lambda_\pm)^{k+1}$ implies, we must write $p_k^N(z) = A_z(\lambda_+)^{k-1} + B_z(\lambda_-)^{k-1}$ and $r_k^N(z) = A_z(\lambda_+)^{k+1} + B_z(\lambda_-)^{k+1}$, where $A_z$ and $B_z$ are to be determined from the boundary conditions. Upon determining $A_z$ and $B_z$ from the boundary conditions, we have $A_z = \frac{z(\lambda_-)^N}{(\lambda_-)^N - (\lambda_+)^N}$ and $B_z = \frac{z(\lambda_+)^N}{(\lambda_+)^N - (\lambda_-)^N}$, and the lemma follows.

Now take $a = b = c = -d = \frac{1}{\sqrt{2}}$ in Lemma 1 of Konno et al. [3], and apply Fourier inversion, see [2] for example, then the correction for the finite Hadamard walk is the following:

**Theorem 2.2.2.** $P_k^N(\varphi) = c_1|\alpha|^2 + c_2|\beta|^2 + 2\operatorname{Re}(c_3\overline{\alpha}\beta)$, *where* $\varphi =^t [\alpha, \beta] \in \Phi$ *and*

$$c_1 = \frac{1}{2\pi}\int_0^{2\pi} \left|\frac{1}{\sqrt{2}}p_k^N(e^{i\theta}) + \frac{1}{\sqrt{2}}r_k^N(e^{i\theta})\right|^2 d\theta,$$

$$c_2 = \frac{1}{2\pi}\int_0^{2\pi} \left|\frac{1}{\sqrt{2}}p_k^N(e^{i\theta}) - \frac{1}{\sqrt{2}}r_k^N(e^{i\theta})\right|^2 d\theta,$$

$$c_3 = \frac{1}{2\pi}\int_0^{2\pi} \left(\overline{\frac{1}{\sqrt{2}}p_k^N(e^{i\theta}) + \frac{1}{\sqrt{2}}r_k^N(e^{i\theta})}\right)\left(\frac{1}{\sqrt{2}}p_k^N(e^{i\theta}) - \frac{1}{\sqrt{2}}r_k^N(e^{i\theta})\right) d\theta,$$

*where* $p_k^N(z)$ *and* $r_k^N(z)$ *satisfy Lemma* 2.2.1, *and the bar denotes complex conjugation.*

Let us remark in the case of the finite Hadamard walk that Theorem 2.2.2 could be interpreted as the absorption probability for the particle irrespective of the initial location $k \in \{1, 2, ..., N-1\}$ and the initial qubit state $\varphi \in \Phi$.

The particular interest of the authors in [3] is the case where $k = 1$, $\varphi =^t [0, 1] = |R\rangle$. In this case, we have from Theorem 2.2.2 the following:

**Corollary 2.2.3.** *For* $N \geq 2$,

$$P_1^N(^t[0, 1]) = \frac{1}{2}\left(1 + \frac{1}{2\pi}\int_0^{2\pi} |r_1^N(e^{i\theta})|^2 d\theta\right),$$

*where* $r_1^N(z)$ *satisfies*

$$r_1^N(z) = \frac{z(\lambda_-)^N}{(\lambda_-)^N - (\lambda_+)^N}\lambda_+^2 + \frac{z(\lambda_+)^N}{(\lambda_+)^N - (\lambda_-)^N}\lambda_-^2.$$

**Proof.** To see $r_1^N(z)$ take $k = 1$ in the expression for $r_k^N(z)$ in Lemma 2.2.1. Now notice that $r_1^N(z)$ is odd, thus, $\overline{r_1^N(z)} = r_1^N(\bar{z})$, where the bar denotes complex conjugation. The authors in [3] have noted $p_1^N(z) = z$ for $N \geq 2$. Since $p_1^N(z) = z$ is also odd, $p_1^N(\bar{z}) = \overline{p_1^N(z)}$. So $c_1$, $c_2$ and $c_3$ in Theorem 2.2.2 can be written, respectively, as

$$c_1 = \frac{1}{2\pi}\int_0^{2\pi}\left(\frac{1}{\sqrt{2}}e^{i\theta} + \frac{1}{\sqrt{2}}r_1^N(e^{i\theta})\right)\left(\frac{1}{\sqrt{2}}e^{-i\theta} + \frac{1}{\sqrt{2}}r_1^N(e^{-i\theta})\right)d\theta, \quad (2.2.4)$$

$$c_2 = \frac{1}{2\pi}\int_0^{2\pi}\left(\frac{1}{\sqrt{2}}e^{i\theta} - \frac{1}{\sqrt{2}}r_1^N(e^{i\theta})\right)\left(\frac{1}{\sqrt{2}}e^{-i\theta} - \frac{1}{\sqrt{2}}r_1^N(e^{-i\theta})\right)d\theta, \quad (2.2.5)$$

$$c_3 = \frac{1}{2\pi}\int_0^{2\pi}\left(\frac{1}{\sqrt{2}}e^{-i\theta} + \frac{1}{\sqrt{2}}r_1^N(e^{-i\theta})\right)\left(\frac{1}{\sqrt{2}}e^{i\theta} - \frac{1}{\sqrt{2}}r_1^N(e^{i\theta})\right)d\theta. \quad (2.2.6)$$

Now simplifying the integrands in (2.2.4), (2.2.5) and (2.2.6), we see that we can write

$$c_1 = \frac{1}{2} + \frac{1}{4\pi}\int_0^{2\pi} e^{i\theta} r_1^N(e^{-i\theta}) d\theta + \frac{1}{4\pi}\int_0^{2\pi} e^{-i\theta} r_1^N(e^{i\theta}) d\theta$$

$$+ \frac{1}{4\pi}\int_0^{2\pi} |r_1^N(e^{i\theta})|^2 d\theta, \quad (2.2.7)$$

$$c_2 = \frac{1}{2} - \frac{1}{4\pi}\int_0^{2\pi} e^{i\theta} r_1^N(e^{-i\theta})d\theta - \frac{1}{4\pi}\int_0^{2\pi} e^{-i\theta} r_1^N(e^{i\theta})d\theta$$

$$+ \frac{1}{4\pi}\int_0^{2\pi} |r_1^N(e^{i\theta})|^2 d\theta, \qquad (2.2.8)$$

$$c_3 = \frac{1}{2} - \frac{1}{4\pi}\int_0^{2\pi} e^{-i\theta} r_1^N(e^{i\theta})d\theta + \frac{1}{4\pi}\int_0^{2\pi} e^{i\theta} r_1^N(e^{-i\theta})d\theta$$

$$- \frac{1}{4\pi}\int_0^{2\pi} |r_1^N(e^{i\theta})|^2 d\theta. \qquad (2.2.9)$$

Let $u = e^{-i\theta}$. Then changing the variable of integration in $\int_0^{2\pi} e^{-i\theta} r_1^N(e^{i\theta})d\theta$ and

$\int_0^{2\pi} e^{i\theta} r_1^N(e^{-i\theta})d\theta$, we have

$$\int_0^{2\pi} e^{-i\theta} r_1^N(e^{i\theta})d\theta = \frac{1}{-i}\int_1^1 r_1^N\left(\frac{1}{u}\right)du = 0$$

and

$$\int_0^{2\pi} e^{i\theta} r_1^N(e^{-i\theta})d\theta = \frac{1}{-i}\int_1^1 \frac{1}{u} r_1^N(u)du = 0.$$

Using this in (2.2.7), (2.2.8) and (2.2.9), we have

$$c_1 = c_2 = \frac{1}{2}\left[1 + \frac{1}{2\pi}\int_0^{2\pi} |r_1^N(e^{i\theta})|^2 d\theta\right], \qquad (2.2.10)$$

$$c_3 = \frac{1}{2}\left[1 - \frac{1}{2\pi}\int_0^{2\pi} |r_1^N(e^{i\theta})|^2 d\theta\right]. \qquad (2.2.11)$$

Since $|\alpha|^2 + |\beta|^2 = 1$, using (2.2.10) and (2.2.11) in $P_k^N(\varphi) = c_1|\alpha|^2 + c_2|\beta|^2 + 2\operatorname{Re}(c_3\bar{\alpha}\beta)$, we get

$$P_1^N(\varphi) = \frac{1}{2}\left[1 + \frac{1}{2\pi}\int_0^{2\pi} |r_1^N(e^{i\theta})|^2 d\theta\right]$$

$$+ \frac{1}{2}\left[1 - \frac{1}{2\pi}\int_0^{2\pi} |r_1^N(e^{i\theta})|^2 d\theta\right][2\operatorname{Re}(\bar{\alpha}\beta)]. \qquad (2.2.12)$$

The corollary follows by letting $\varphi =^t [0, 1]$ in (2.2.12).

### 3. The Falsity of the Conjecture

If $N = 3$, then we have from Corollary 2.2.3,

$$P_1^3(\,^t[0, 1]) = \frac{1}{2}\left(1 + \frac{1}{2\pi}\int_0^{2\pi} |\,r_1^3(e^{i\theta})\,|^2 d\theta\right),$$

where

$$r_1^3(z) = \frac{z(\lambda_-)^3}{(\lambda_-)^3 - (\lambda_+)^3}\lambda_+^2 + \frac{z(\lambda_+)^3}{(\lambda_+)^3 - (\lambda_-)^3}\lambda_-^2.$$

The authors in [3] have shown that

$$\lambda_\pm = \frac{(z^2 - 1) \pm \sqrt{z^4 + 1}}{\sqrt{2}z}, \quad \lambda_+\lambda_- = -1, \quad \lambda_+ + \lambda_- = \sqrt{2}\left(z - \frac{1}{z}\right).$$

Using these we can write $r_1^3(z) = \dfrac{z^3}{2z^4 - 3z^2 + 2}$. It is easily seen that $r_1^3(z)$ is odd, thus, $r_1^3(\bar{z}) = \overline{r_1^3(z)}$. So we can write

$$P_1^3(\,^t[0, 1]) = \frac{1}{2}\left(1 + \frac{1}{2\pi}\int_0^{2\pi} r_1^3(e^{i\theta})r_1^3(e^{-i\theta})d\theta\right).$$

Write

$$F(\theta) = \int r_1^3(e^{i\theta})r_1^3(e^{-i\theta})d\theta.$$

Using Wolfram Mathematica Online Integrator, we have

$$F(\theta) = \frac{-(-4 + 3e^{2i\theta})}{14(-3e^{2i\theta} + 2e^{4i\theta} + 2)} + \frac{3\log(-4ie^{2i\theta} - 3i + \sqrt{7})}{14\sqrt{7}}$$

$$- \frac{3\log(4ie^{2i\theta} - 3i + \sqrt{7})}{14\sqrt{7}},$$

from which we can show that $F(2\pi) - F(0) = 0$. Thus, $\int_0^{2\pi} r_1^3(e^{i\theta})r_1^3(e^{-i\theta})d\theta = 0$, and so $P_1^3(\,^t[0, 1]) = \frac{1}{2}(1 + 0) = \frac{1}{2}$ which contradicts the assertion made by the conjecture that $P_1^3(\,^t[0, 1]) = \frac{2}{3}$.

ON THE AMBAINIS-BACH-NAYAK-VISHWANATH-WATROUS ...    9## 4. Concluding Remarks and Open Problems

If $U = H$ (the Hadamard walk), then we have given a formula for $P_k^N(\varphi)$ irrespective of the initial location $k \in \{1, ..., N-1\}$ and the initial quibit state $\varphi \in \Phi$ of the particle. In the particular case, $k = 1$ and $\varphi =^t [0, 1] = |R\rangle$, we have shown Conjecture 11 of [3] does not hold in the case $N = 3$. If $U \neq H$, then there still remains the problem of finding a formula for $P_k^N(\varphi)$ in the likeness of Theorem 2.2.2 of this paper. This problem has arisen as we have shown the flaw in Theorem 2 of [3]. A more interesting problem will be to find absorption probabilities for quantum walks in general dimensionality with or without "traps."

## References

[1] A. Ambainis, E. Bach, A. Nayak, A. Vishwanath and A. Watrous, One-dimensional Quantum Walks, Proceedings of the 33rd Annual ACM Symposium on the Theory of Computing (2001), 37-49.

[2] G. Folland, Fourier Analysis and its Applications, Wadsworth & Brooks/Cole Advanced Books and Software, Pacific Grove, California, 1992.

[3] N. Konno, T. Namiki, T. Soshi and A. Sudbury, Absorption problems for quantum walks in one dimension, J. Phys. A: Math. Gen. 36 (2003), 241-253.